\documentclass[aps,prl,twocolumn,groupedaddress,draft]{revtex4-2}

\usepackage{amsmath}
\usepackage{amssymb}
\usepackage{color}

\newcommand{\nc}{\newcommand}

\newcommand{\half}{\tfrac{1}{2}}
\def\Tr{{\mathbf{Tr}}}

\nc{\ms}{M_S^2}

\nc{\hsp}{\hspace{0.3cm}}
\nc{\lsp}{\hspace{1cm}}
\nc{\non}{\nonumber}
\nc{\dd}{\partial}

\newcommand{\beq}{\begin{equation}}  \newcommand{\eeq}{\end{equation}}
\newcommand{\bea}{\begin{eqnarray}}  \newcommand{\eea}{\end{eqnarray}}
\newcommand{\baa}{\begin{array}}     \newcommand{\eaa}{\end{array}}
\newcommand{\bit}{\begin{itemize}}   \newcommand{\eit}{\end{itemize}}
\newcommand{\ben}{\begin{enumerate}} \newcommand{\een}{\end{enumerate}}
\newcommand{\bce}{\begin{center}}    \newcommand{\ece}{\end{center}}
\newcommand{\bpm}{\begin{pmatrix}}   \newcommand{\epm}{\end{pmatrix}}
\newcommand{\bvt}{\begin{verbatim}}  \newcommand{\evt}{\end{verbatim}}

\begin{document}


\title{Imaginary scaling invariance of the one-loop effective potential}

\author{P. M. Ferreira}
\email[]{pmmferreira@fc.ul.pt}
\affiliation{Instituto  Superior  de  Engenharia  de  Lisboa,  Portugal}
\affiliation{Centro  de  F{\'i}sica  Te{\'o}rica  e  Computacional,  Universidade  de  Lisboa,  Portugal}

\author{B. Grzadkowski}
\email[]{bohdan.grzadkowski@fuw.edu.pl}
\affiliation{Faculty of Physics, University of Warsaw, Pasteura 5, 02-093 Warsaw, Poland}

\author{O. M. Ogreid}
\email[]{omo@hvl.no}
\affiliation{Western Norway University of Applied Sciences, Postboks 7030, N-5020 Bergen, Norway}


\date{\today}

\begin{abstract}
Recently, a hitherto unknown class of renormalization group stable relations between parameters of bosonic field
theories has been identified and dubbed as the $r_0$ or ``GOOFy" symmetries. 
Here, one-loop properties of the $r_0$ invariant two-Higgs doublet model and a minimal symmetric model are discussed.
It is concluded that the symmetry is present at the one-loop provided the UV cutoff squared transforms
nontrivially under $r_0$. The minimal model requires the presence of two real fields.
\end{abstract}


\maketitle

\section{Introduction}
Investigating symmetries of the two-Higgs doublet odel (2HDM) extension of the Standard Model (SM) of electro-weak interactions,
the authors of Ref.~\cite{Ferreira:2023dke} have recently  noticed the existence of a new class of relations between 
parameters of
the 2HDM scalar potential which are stable under the renormalization group equations (RGE)up to an infinite order of 
perturbation expansion.
It turned out the relations are not implied by hitherto known symmetries of the 2HDM, so a new class of symmetries 
has been hypothesized
in Ref.~\cite{Ferreira:2023dke}. The symmetry transformations, dubbed as the $r_0$ or GOOFy\footnote{The acronym GOOFy consists of the first letters in the surnames of the authors that suggested this symmetry; Grzadkowski, Ogreid, Osland, Ferreira} symmetries,
include mapping real bosonic fields (scalars and vectors) into purely imaginary ones.
Even if the tree-level potential and Lagrangian density have been shown to be invariant under the $r_0$ transformation, the invariance could be broken at higher orders in perturbation theory.
In order to investigate if the new transformations constitute
genuine symmetries, we shall investigate properties of the one-loop effective potential of the 2HDM. In addition, we introduce, present and discuss in detail a minimal $r_0$-invariant scalar field model~\cite{Boh:2024MPI}.

The idea of GOOFy symmetries has been picked up by Haber and Ferreira and applied to complexified GOOFy-symmetric models in Ref.~\cite{Haber:2025cbb}. Then Trautner developed the idea of GOOFy symmetry for generic scalar field theories in \cite{Trautner:2025yxz}, by considering noninvariance of kinetic terms.
The new symmetries could be naturally applied to the hierarchy problem; that direction has been picked up in Refs.~\cite{deBoer:2025jhc,Trautner:2025prm}.
Interesting discussions of the $r_0$ symmetry at the one-loop level has recently appeared in Ref.~\cite{Pilaftsis:2024uub}.

It should be noted that using the term ``symmetries" to describe these findings is still open to debate. On the one hand, 
the procedure
discovered in~\cite{Ferreira:2023dke} has allowed the identification of hitherto unknown RGE-protected, to all orders,
relations between parameters, 
which is certainly the hallmark of a symmetry. On the other hand, the process that seems to lead to those parameter 
regions involves imaginary
scalings of both real components of scalar doublets (and gauge fields) and spacetime coordinates, which is unlike any
symmetry used
in quantum field theory. In  \cite{Trautner:2025yxz} a possible interpretation involving spurions was proposed, and 
in~\cite{Haber:2025cbb}  RGE-stable relations found through the GOOFy paradigm in a two real scalar model were interpreted
as the consequence of a regular symmetry imposed on a model with twice the field content. Nonetheless, we argue that the
GOOFy procedure has been shown to produce physical consequences, such as mass degeneracies or parameter RGE stability, for
a growing number of different models. And in this work, we will show that invariance under the GOOFy transformations is
not only found at tree level but also for the one-loop effective potential. We therefore argue that indeed the epithet
``symmetries" seems to apply to the field transformations herein discussed, no matter how bizarre they may seem.

\section{The Two-Higgs Doublet Model}
\label{sec:2hdm}

The 2HDM is one of the simplest extensions of the SM, wherein one considers two $SU(2)$ doublets with hypercharge one
instead of just one doublet. In the following we will briefly review the basic aspects of a useful formalism
to understand the structure of the scalar sector of the model, and the global symmetries one can impose
upon it.
The most general scalar potential involving two hypercharge $Y = 1$ scalar doublets invariant
under the electroweak gauge group $SU(2)_L\times U(1)_Y$ is given by
\begin{widetext}
\bea
V_\text{tree} &=& m_{11}^2\Phi_1^\dagger\Phi_1+m_{22}^2\Phi_2^\dagger\Phi_2
-[m_{12}^2\Phi_1^\dagger\Phi_2+{\rm h.c.}]+\half\lambda_1(\Phi_1^\dagger\Phi_1)^2
+\half\lambda_2(\Phi_2^\dagger\Phi_2)^2
+\lambda_3(\Phi_1^\dagger\Phi_1)(\Phi_2^\dagger\Phi_2)\nonumber\\[8pt]
&&\quad
+\lambda_4(\Phi_1^\dagger\Phi_2)(\Phi_2^\dagger\Phi_1)
+\left\{\half\lambda_5(\Phi_1^\dagger\Phi_2)^2
+\big[\lambda_6(\Phi_1^\dagger\Phi_1)
+\lambda_7(\Phi_2^\dagger\Phi_2)\big]
\Phi_1^\dagger\Phi_2+{\rm h.c.}\right\}\,,
\label{eq:pot}
\eea
\end{widetext}
where, other than $m^2_{12}$ and $\lambda_{5,6,7}$, all parameters are real. An alternative notation uses
four gauge-invariant bilinears constructed from the
doublets~\cite{Velhinho:1994np,Nagel:2004sw,Ivanov:2005hg,Ivanov:2006yq,Ivanov:2007de,Maniatis:2006fs,Maniatis:2006jd,
	Nishi:2006tg,Nishi:2007nh,Nishi:2007dv,Maniatis:2007vn},
\beq
\begin{array}{rcl}
	r_0 &=&
	\frac{1}{2}
	\left( \Phi_1^\dagger \Phi_1 + \Phi_2^\dagger \Phi_2 \right),
	\\*[2mm]
	r_1 &=&
	\frac{1}{2}
	\left( \Phi_1^\dagger \Phi_2 + \Phi_2^\dagger \Phi_1 \right)
	= \mbox{Re}\left( \Phi_1^\dagger \Phi_2 \right),
	\\*[2mm]
	r_2 &=&
	- \frac{i}{2}
	\left( \Phi_1^\dagger \Phi_2 - \Phi_2^\dagger \Phi_1 \right)
	= \mbox{Im} \left( \Phi_1^\dagger \Phi_2 \right),
	\\*[2mm]
	r_3 &=&
	\frac{1}{2}
	\left( \Phi_1^\dagger \Phi_1 - \Phi_2^\dagger \Phi_2 \right).

\end{array}
\label{eq:rs}
\eeq
In terms of these quantities, then, the potential of Eq.~\eqref{eq:pot} may be written as
\beq
V \,=\,M_\mu\,r^\mu\,+\,\Lambda_{\mu\nu}\,r^\mu\,r^\nu\,,
\label{potbi}
\eeq
where we use a Minkowski-like formalism to define the four-vectors
\bea
r^\mu &=& (r_0\,,\,r_1\,,\,r_2\,,\,r_3) \,=\, (r_0\,,\,\vec{r})\,, \nonumber \\
M^\mu &=& \left(m^2_{11} + m^2_{22}\,,\, 2\mbox{Re}(m^2_{12})
\,,\, -2\mbox{Im}(m^2_{12})\,,\,m^2_{22} - m^2_{11}\right)\nonumber\\
&=&\, (M_0\,,\,\vec{M})\,,
\label{eq:defbil}
\eea
as well as the tensor
\begin{widetext}
\begin{align}
	\Lambda^{\mu\nu} & = \begin{pmatrix} \Lambda_{00} & \vec{\Lambda} \\
		\vec{\Lambda}^T & \Lambda \end{pmatrix}
	=
	\begin{pmatrix}
		\frac{1}{2}(\lambda_1 + \lambda_2) + \lambda_3 &
		-\mbox{Re}\left(\lambda_6 + \lambda_7\right) &
		\mbox{Im}\left(\lambda_6 + \lambda_7\right) &
		\frac{1}{2}(\lambda_2 - \lambda_1) \\
		-\mbox{Re}\left(\lambda_6 + \lambda_7\right) &
		\lambda_4 + \mbox{Re} \left( \lambda_5\right) &
		- \mbox{Im} \left( \lambda_5 \right) &
		\mbox{Re}\left(\lambda_6 - \lambda_7\right)
		\\
		\mbox{Im} \left( \lambda_6 + \lambda_7\right)&
		- \mbox{Im} \left( \lambda_5 \right) &
		\lambda_4 - \mbox{Re} \left( \lambda_5\right) &
		- \mbox{Im} \left( \lambda_6 - \lambda_7\right) \\
		\frac{1}{2}(\lambda_2 - \lambda_1) &
		\mbox{Re}\left( \lambda_6 - \lambda_7\right) &
		- \mbox{Im} \left( \lambda_6 - \lambda_7 \right) &
		\frac{1}{2}(\lambda_1 + \lambda_2) - \lambda_3
	\end{pmatrix}\,.
	\label{eq:Lambda}
\end{align}
\end{widetext}

In our previous publication~\cite{Ferreira:2023dke} it was shown that there exist hitherto
unknown relations between parameters of the scalar potential which are stable under RGE to all orders in the
scalar and gauge sectors, and at least up to two loops when the Yukawa sector is included\footnote{It would be interesting and relevant to incorporate fermions into this discussion  as well. However, in order to preserve the letter format, fermions are relegated to future projects, which are, in fact, in preparation.}. 
In other words, the corresponding beta functions have fixed points.
In~\cite{Ferreira:2023dke} we hypothesized the existence of a class of new symmetries, called $r_0$ symmetries 
for the 2HDM since they seem to arise
from the transformation $r_0\rightarrow -r_0$, which would explain the presence of those relations.
It should be mentioned here that according to common knowledge, there exist only six possible symmetries of the 2HDM scalar 
potential~\cite{Ivanov:2005hg,Ivanov:2006yq} that are consistent with gauge $SU(2)\times U(1)$ invariance. This work states that
there exist six extra symmetries that adopt the $r_0$ transformation~\cite{Ferreira:2023dke}. 
Phenomenology of the 2HDM with the $r_0$ symmetries has been presented in Sec.~4 of our earlier work~\cite{Ferreira:2023dke},
where we showed that these symmetries could be used to explain mass degeneracies (exact or approximate) in the scalar sector, 
and were also responsible for specific tree-level relations between masses and couplings.
Furthermore, it was also shown that for GOOFy 2HDMs the decoupling of heavy degrees of freedom is not possible,
so that the masses of the extra scalars (other than the SM Higgs) are necessarily below $\sim $ 700 GeV. 
Further details of phenomenology of GOOFy 2HDMs, involving the fermionic sector as well, will be presented 
in~\cite{Ferreira:2026new}, and an extension of GOOFy symmetries to three-Higgs doublet models
is being investigated~\cite{SARA:2026new}.

\section{The $r_0$ symmetry and its consequences}
\label{sec:news}

In the previous publication \cite{Ferreira:2023dke} we identified the following relations between potential parameters of the 2HDM that are fixed points of the renormalization group equations valid to all orders of perturbation theory in the scalar and gauge sectors:
\beq
\vec{\Lambda} = \vec{0} \hsp \text{and} \hsp M_0 = 0\,.
\label{fp}
\eeq
More explicitly, beta functions corresponding to
$m_{11}^2+m_{22}^2$, $\lambda_1-\lambda_2$ and $\lambda_6+\lambda_7$ vanish at the fixed point \eqref{fp}. In terms of parameters of the potential \eqref{eq:pot} the above relations read as $m_{11}^2+m_{22}^2=0$, $\lambda_1=\lambda_2$ and $\lambda_6=-\lambda_7$.

Let us define the following basis in eight-dimensional space for the real components of the two doublets,
\beq
\phi_i \equiv  \left\{ \phi_1,\phi_2,\phi_3,\phi_4,\phi_5,\phi_6, \phi_7,\phi_8 \right\}
\label{field_basis}
\eeq
such that
\bea
\Phi_1=\frac{1}{\sqrt{2}}
\begin{pmatrix}
	\phi_1+i\phi_2 \\
	\phi_3+i\phi_4
\end{pmatrix}, \quad
\Phi_2=\frac{1}{\sqrt{2}}
\begin{pmatrix}
	\phi_5+i\phi_6 \\
	\phi_7+i\phi_8
\end{pmatrix}\,.
\eea
Given that $r_0$ of Eq. (\ref{eq:rs}) is obviously non-negative, it is impossible to change the sign of $r_0$ without resorting to transforming real fields into purely imaginary fields.
It is easy to verify that the field transformations of the eight real components of $\Phi_{1,2}$ that imply \eqref{fp} are the following:
\bea
	&&\phi_1\to i\phi_6, \hsp \phi_2\to i\phi_5, \hsp \phi_3\to i\phi_8, \hsp \phi_4\to i\phi_7, \label{r0_trans} \\
	&&\phi_5\to -i\phi_2, \hsp \phi_6\to -i\phi_1, \hsp \phi_7\to -i\phi_4, \hsp \phi_8\to -i\phi_3\,. \nonumber
\eea
The above transformations imply $r_0\rightarrow - r_0$, as one could have anticipated.
It has been also shown in \cite{Ferreira:2023dke} that in order to keep scalar and gauge kinetic terms invariant under the above transformation, the additional transformations of the coordinate $x^\mu$ and vector fields are needed:
\beq
\begin{split}
	&x^\mu \to i x^\mu, \quad
	B_\mu \to i B_\mu, \\
	&W_{1\mu} \to i W_{1\mu},\quad
	W_{2\mu} \to -i W_{2\mu},\quad
	W_{3\mu} \to i W_{3\mu}\,.
\end{split}
\label{eq:igau}
\eeq
Hereafter we will refer to \eqref{r0_trans} together with \eqref{eq:igau} as the $r_0$ transformations.
Note that, by construction, the Lagrangian density ${\cal L}$ is invariant under the $r_0$ transformation, 
and it is easily shown that the classical action is invariant as well.


Let us now consider a more general system that consists of a number of real scalar fields in order to find out what is the scaling for the momenta  of the system that is implied by a given scaling of fields and coordinates. Relativistic quantum mechanics suggests that $x^\mu \to i x^\mu$ implies  $p_\mu\equiv -i\partial_\mu  \to -\,i\,p_\mu$. In fact, in field theory the 4-momentum is given in terms of the energy-momentum tensor $\Theta_{\alpha\beta}$ as
\beq
p_\mu \,\equiv \,\int \,d^3 x\, \Theta_{0\mu}\,,
\label{eq:4P}
\eeq
with
\beq
\Theta_{\alpha\beta}\,=\, \sum_i \,\frac{\partial \cal L}{\partial (\partial^\alpha \phi_i)}\,\partial_\beta\,\phi_i\;- \;
\eta_{\alpha\beta}\,\cal L\,.
\label{eq:Theta}
\eeq
The Lagrangian ${\cal L}$ is invariant under the $r_0$ transformation
(\ref{r0_trans}) and (\ref{eq:igau}), therefore $\Theta_{\alpha\beta}$ is invariant as well. On the other hand
$d^3 x \to -i d^3 x$; therefore, taking into account integration limits, we confirm that indeed $p_\mu \to -\,i\,p_\mu$, in agreement with our initial intuition.



\section{Invariance of the 2HDM effective potential under the $r_0$ transformation}

In order to test the invariance of the effective potential under the $r_0$ transformation, it is convenient to invoke its functional
definition~\cite{Jackiw:1974cv, Itzykson:1980rh, Ramond:1981pw}, since it allows the formulation of the potential in terms of coordinates and fields
only. Adopting, e.g.,~\cite{Itzykson:1980rh} one finds that the one-loop contribution to the effective potential in a generic real scalar model reads as
\begin{widetext}
\beq
V_{\text{eff}}(\phi_{cl}) \propto \int \prod_k {\cal D}(\phi_k) \exp \left\{ -i\int d^4x \; \phi_i(x) \left[ \Box_x \; \delta_{ij} + \left(M^2_S(\phi_{cl})\right)_{ij}  \right] \phi_j(x) \right\},
\label{V_func}
\eeq
\end{widetext}
where $\Box_x \equiv \partial_\mu\partial^\mu$ and $M^2_S(\varphi_{cl})$ is the scalar mass 
squared matrix,
\beq
\left(M_S^2\right)_{ij} \equiv \frac{\partial^2 V_\text{tree}}{\partial\phi_i\partial\phi_j}\,
\label{eq:MS2}
\eeq
calculated for a given tree-level potential for constant classical field configuration $\phi_{cl}$.
The d'Alembertian is odd under $r_0$, i.e. $\Box_x \to -\Box_x$. Since $\phi_i \to \pm i \phi_i$, it becomes clear that, for both models 
considered in this paper, the measure
$\prod_k{\cal D}(\phi_k)$ is invariant under the  $r_0$ transformation, so we conclude that the one-loop effective potential is indeed $r_0$ invariant.

Recently, within the 2HDM, the one-loop effective potential has been calculated adopting the bilinear
notation~\cite{Cao:2022rgh,Cao:2023kgq,Pilaftsis:2024uub}.
We will concentrate hereafter only on the scalar ($V_{\text{eff}}^{(S)}$) contribution to the potential that can be,
after dropping field-independent contributions, written in the following form \cite{Quiros:1999jp}
\begin{align}
	V_{\text{eff}}^{(S)} &=\frac{1}{2}\int
	\frac{d^4p_E}{(2\pi)^4}\mathbf{Tr}\left[\ln (p_E^2+M_S^2)\right]\nonumber\\
	&= -\frac{1}{2}\int
	\frac{d^4p_E}{(2\pi)^4}\left[\mathbf{Tr}\sum_{n=1}^\infty\frac{(-1)^n}{n}\left(\frac{M_S^2}{p^2_E}\right)^n\right]\,,
	\label{eq:VCW}
\end{align}
where $p_E$ is the Euclidian momentum obtained by a standard Wick rotation from the 4-momentum of virtual quanta propagating in a loop.
At the lowest order, $n=1$, specializing to the 2HDM fixed point \eqref{fp}, and calculating the trace in the $i,j$ indices, we obtain
the $r_0$-odd result
\beq
\mathbf{Tr}(M_S^2)=
4[5\Lambda_{00}+\mathbf{tr}(\Lambda)]r_0.
\label{mass_pot3}
\eeq
In order to verify the transformation properties of higher order terms let us denote by $\lambda_i(r_0)$ an eigenvalue of $M_S^2(r_0)$. For the 2HDM,
the characteristic equation for $M_S^2$ could be written as follows 
\beq
\lambda^8 + c_7(r_0)\lambda^7+c_6(r_0)\lambda^6+\cdots+ c_0(r_0)=0,
\label{char_eq1}
\eeq
where $c_i(r_0)$ are coefficients determined in terms of entries of the matrix $M_S^2$, in particular $c_0(r_0)=\det\left(M_S^2\right)$ and $c_7(r_0)=\text{Tr}\left(M_S^2\right)$.
It turns out that the coefficients satisfy the following symmetry relations:
\beq
c_{2n}(r_0)=c_{2n}(-r_0)\;\text{and}\; c_{2n+1}(r_0)=-c_{2n+1}(-r_0)\,.
\label{sym_rel}
\eeq
Writing the characteristic equation for $M_S^2(-r_0)$
\beq
\lambda^8 + c_7(-r_0)\lambda^7+c_6(-r_0)\lambda^6+\cdots+ c_0(-r_0)=0,
\label{char_eq2}
\eeq
and adopting \eqref{sym_rel} one can conclude that if $\lambda_i(r_0)$ is an eigenvalue of $M_S^2(r_0)$ then $-\lambda_i(r_0)$ 
is an eigenvalue of $M_S^2(-r_0)$, i.e. there exists an eigenvalue $\lambda_j(r_0)$ such that
\bea
&&\lambda_i(r_0) \stackrel{r_0}{\longrightarrow} \lambda_i(-r_0)=-\lambda_j(r_0),
\hsp \text{and} \nonumber\\
&&\lambda_j(r_0) \stackrel{r_0}{\longrightarrow} \lambda_j(-r_0)=-\lambda_i(r_0).
\label{eigen_trans_2}
\eea
So there are four pairs of eigenvalues transforming within each pair, so that the set of the eigenvalues $\{\lambda_i(r_0)\}$ transforms as follows:
\beq
\{\lambda_i(r_0)\}\stackrel{r_0}{\longrightarrow} -\{\lambda_i(r_0)\}.
\label{lam_trans}
\eeq
An important conclusion from the above is that
\bea
\mathbf{Tr}\left[\left(M_S^2\right)^{2n}\right](r_0)&=& \mathbf{Tr}\left[\left(M_S^2\right)^{2n}\right](-r_0), \;\;\text{and}\nonumber\\ \mathbf{Tr}\left[\left(M_S^2\right)^{2n+1}\right](r_0)&=&-\mathbf{Tr}\left[\left(M_S^2\right)^{2n+1}\right](-r_0).
\label{trace_par}
\eea
For any finite $n$ the trace of higher powers of $\ms$ could be calculated explicitly.
Indeed, considering the definition of $M_S^2$, Eq.~\eqref{eq:MS2}, it is simple to verify Eq.~\eqref{trace_par} for all values of $n$.
Therefore we see that terms with even/odd powers of
$(\ms/p_E^2)$ appearing in \eqref{eq:VCW} are even/odd functions of $r_0$.
But since under the $r_0$ transformation we must have $p_E^2 \to -p_E^2$ (a consequence of $p_\mu \to -\,i\,p_\mu$) this results in the $r_0$ invariance of the
effective potential, equivalent to the $r_0$-invariance of (\ref{V_func}).

In order to complete the calculation of the effective potential \eqref{eq:VCW} one needs to perform the integration over $d^4p_E$, where some regularization strategy must be adopted. For instance, the cutoff regularization implies that
\beq
V_{\text{eff}}^{(S)} = -\frac{1}{32\pi^2} \sum_{n=1}^\infty\left[ \frac{(-1)^n}{n}\int_0^{\Lambda_{UV}^2}
d\rho \rho \; \mathbf{Tr} \left(\frac{M_S^2}{\rho}\right)^n\right]\,,
\label{eq:VCW_cut}
\eeq
where $\rho\equiv p_E^2$. Note that if we replace $\Lambda_{UV}^2$ by $-\Lambda_{UV}^2$ this is equivalent to replacing $\rho$ by $-\rho$ under the trace.
Consequently, for odd $n$ a minus sign appears. Since for odd $n$ $\mathbf{Tr} \left(M_S^2(r_0)\right)^n$ is an odd function of $r_0$, but under the $r_0$
transformation we have $\Lambda_{UV}^2\to-\Lambda_{UV}^2$, we hence arrive at an invariant one-loop effective potential~\footnote{If we had adopted dimensional
	regularization, the regularization scale $\mu^2$ would be found to be odd under the $r_0$ transformation: $\mu^2\to -\mu^2$.}.
This is indeed consistent with our earlier conclusion that under $r_0$, $p_\mu \to -\,i\,p_\mu$.

The $r_0$ invariance could be conveniently checked in the basis in which $M_S^2$ is diagonal: $(M_S^2)_{ij}=\delta_{ij}\lambda_j=\delta_{ij}M_j^2$,
where there is no summation over $j$. Adopting the cutoff regularization, neglecting field independent terms
and terms which vanish in the limit $\Lambda_{UV} \to \infty$, one finds the effective potential
\bea
V_{\text{eff}}^{\text{1-loop}}(r_0)&=&
\frac{\Lambda_{\text{UV}}^2}{32\pi^2} \sum_{i=1}^8 M_i^2(r_0) \label{eff_pot_cut} \\
&&+
\frac{1}{64\pi^2} \sum_{i=1}^8 M_i^4(r_0)\left[\log\frac{M_i^2(r_0)}{\Lambda_{\text{UV}}^2}-\frac{1}{2} \right]\,.\nonumber
\eea
Using the result \eqref{lam_trans} it is clear that the first term above is invariant since $\Lambda_{UV}^2 \to -\Lambda_{UV}^2$.
On the other hand \eqref{lam_trans} implies that the same transformation of the cutoff guarantees invariance of the second , thus proving 
the invariance of the one-loop effective potential under the $r_0$ transformation.

Note that we have proven invariance without needing to diagonalize the $8\times 8$ mass matrix $M_S^2$. It was sufficient to find out how 
the eigenvalues of $M_S^2$ transform under $r_0$.

\section{The toy model--2RSM}

Even though we have managed, for the 2HDM, to verify transformation properties of the
effective potential without calculating the $M_S^2$ eigenvalues, hereafter we are going to present a simple, minimal toy model~\cite{Boh:2024MPI}
which shares many properties with the 2HDM and for which the diagonalization could be done explicitly. This is a two real scalar model described by
the Lagrangian
\beq
{\cal{L}}= \frac12(\partial_\mu\phi_1\partial^\mu\phi_1 + \partial_\mu\phi_2\partial^\mu\phi_2)
-V(\phi_1,\phi_2)\,,
\label{lag_toy2}
\eeq
with the potential
\bea
V^0(\phi_1,\,\phi_2)&=&\frac12 m_1^2(\phi_1^2-\phi_2^2)+m_{12}^2\phi_1\phi_2\nonumber\\&&
+\frac12 \lambda_1(\phi_1^4+\phi_2^4)+\lambda_3(\phi_1\phi_2)^2\nonumber\\
&&+\lambda_6(\phi_1^2-\phi_2^2)\phi_1\phi_2\,.
\label{pot2}
\eea
The model is invariant under the following $r_0$-like transformation
\beq
x^\mu \to i x^\mu, \lsp \phi_1\to i \phi_2, \lsp \phi_2 \to - i \phi_1\,.
\label{r_0_def}
\eeq
It turns out that for this symmetric potential it is possible to choose a $(\phi_1,\phi_2)$ basis  such that $\lambda_6=0$. This is an analog of the
statement proven for the 2HDM in \cite{Gunion:2005ja,Davidson:2005cw} where the authors showed that if $\lambda_1=\lambda_2$ and
$\lambda_6=-\lambda_7$ ($CP2$ invariant quartic sector) then  there exists a basis such that all $\lambda_i$ are real and $\lambda_6=\lambda_7=0$.
Therefore $\lambda_6$ will be dropped from now on.
The squared mass matrix then reads as
\bea
&&\left(M_S^2\right)_{ij} =\nonumber\\
&&\begin{pmatrix}
	m_1^2+6\lambda_1\phi_1^2+2\lambda_3\phi_2^2 & m_{12}^2+4\lambda_3\phi_1\phi_2 \\
	m_{12}^2+4\lambda_3\phi_1\phi_2 & -m_1^2+6\lambda_1\phi_2^2+2\lambda_3\phi_1^2
\end{pmatrix}\,.
\non
\eea
One can notice that near the origin, $\phi_1=\phi_2=0$, we have
\beq
\left(M_S^2\right)_{ij} =
\begin{pmatrix}
	m_1^2 & m_{12}^2 \\ m_{12}^2 & -m_1^2
\end{pmatrix}
+{\cal{O}}(\phi^2)\,.
\eeq
At the origin the $M_S^2$ eigenvalues are $\pm \sqrt{m_{12}^4+m_1^4}$,
so there is a saddle point at $\phi_1=\phi_2=0$ with two perpendicular directions of opposite curvature (a similar situation appears in the
$r_0$-symmetric 2HDM where, at the origin, the Hessian has four positive and four negative eigenvalues).
This implies that there is a global minimum away from the origin and therefore the $r_0$ symmetry is always spontaneously broken.
It turns out that there are two local minima in the toy model determined by the minimization conditions: 
$(v_1^2-v_2^2) = -m_1^2/\lambda_1$ and $v_1 v_2 = - m_{12}^2/(\lambda_1+\lambda_3)$,
where $\langle \phi_{1,2} \rangle \equiv v_{1,2}/\sqrt{2}$.

One can also express the potential in terms of bilinear variables,
\begin{align}
	r_0 \equiv  \frac12 (\phi_1^2+\phi_2^2)\,,\hsp r_1 \equiv  \phi_1 \phi_2\,, \hsp
	r_2 \equiv  \frac12 (\phi_1^2-\phi_2^2)\,.
\end{align}
Note that $r_0$, $r_1$ and $r_2$ are not independent, since one has $r_0^2-r_1^2-r_2^2=0$.
Upon the $r_0$ transformation,
\beq
(r_0,r_1,r_2) \stackrel{r_0}{\longrightarrow} (-r_0,r_1,r_2)\,.
\eeq
The $r_0$ symmetry thus requires the coefficient of $r_0$ in the potential to vanish, and
the potential may be written as
\beq
V(r^\mu) = -M_\mu r^\mu + \Lambda_{\mu\nu} r^\mu r^\nu\,, \non
\eeq
for $\mu,\nu=0,1,2$ with $r^\mu \equiv (r_0,r_1,r_2)$, $M^\mu\equiv (0,m_{12}^2,m_1^2)$,
the ``Minkowski metric tensor"; $\eta_{\mu\nu}=\text{diag} (1,-1,-1)$; and
\beq
\Lambda^{\mu\nu}\equiv
\begin{pmatrix}
	\Lambda_{00} & 0 & 0 \\ 0 & \Lambda_{11} & \Lambda_{12} \\ 0 & \Lambda_{21} & \Lambda_{22}
\end{pmatrix}
=
\begin{pmatrix}
	\lambda_1 & 0 & 0 \\ 0 & \lambda_3 & 0 \\ 0 & 0 & \lambda_1
\end{pmatrix}\,.
\non
\eeq
Note that $M_0=0$ and $\vec{\Lambda}=0$ are consequences of the $r_0$ symmetry, similar to the 2HDM case, eq.~\eqref{fp}.
Indeed, for the toy model $M_0=0$ and $\vec{\Lambda}=0$ mean that $m_1^2+m_2^2=0$ and $\lambda_1=\lambda_2$.

Then one finds that
\beq
\Tr\left(M_S^2\right)= 4(3\lambda_1+\lambda_3)r^0
\label{trtoy}
\eeq
and therefore under the $r_0$ transformation this trace is odd,
$
\Tr\left(M_S^2\right)  \stackrel{r_0}{\longrightarrow}  -\Tr\left(M_S^2\right)
$,
so that we reproduce the analogous property observed in the 2HDM one-loop effective potential.

Since in this model there are only 2 real degrees of freedom one can easily diagonalize the mass squared matrix and calculate
the one-loop effective potential. It turns out that the eigenvalues of $M_S^2$ can be expressed through bilinears as follows:
\begin{align}
	M_{1,2}^2(r^\mu) & = 2(3\lambda_1+ \lambda_3)r_0 \pm \sqrt{\Delta} \,,
\end{align}
where
\bea
\Delta &=& (m_1^2)^2 + (m_{12}^2)^2 +  4 m_1^2 (3 \lambda_1 - \lambda_3) r_2 + 8 m_{12}^2 \lambda_3 r_1 \nonumber\\
&&+
16 \lambda_3^2 r_0^2 + 12 (3 \lambda_1 + \lambda_3) (\lambda_1 - \lambda_3) r_2^2.
\eea
Note that the eigenvalues of $M_S^2$ are linearly dependent on $r_0$. It is easy to see that they  transform under the $r_0$ symmetry
as
\begin{align}
	M_1^2 \stackrel{r_0}{\longrightarrow}  -M_2^2 \hsp \text{and} \hsp
	M_2^2 \stackrel{r_0}{\longrightarrow}  -M_1^2\,.
\end{align}
In order to illustrate explicitly the $r_0$ invariance at the one-loop level we present the one-loop effective potential adopting cutoff regularization and dropping irrelevant terms:
\bea
V^{\text{1-loop}}_{\text{eff}}(r^\mu)&=&
\frac{\Lambda_{\text{UV}}^2}{32\pi^2} \sum_{i=1,2} M_i^2(r^\mu)\label{eff_pot_cut2}
\\
&&+
\frac{1}{64\pi^2} \sum_{i=1,2}M_i^4(r^\mu)\left[\log\frac{M_i^2(r^\mu)}{\Lambda_{\text{UV}}^2}-\frac12 \right]\,.\nonumber
\eea
Note that
\beq
M_1^4+M_2^4 = 2\left\{[2(3\lambda_1+\lambda_3)r_0]^2 + \Delta\right\}\,,
\eeq
so this quantity is $r_0$ invariant. Here one can explicitly see that the one-loop effective potential is indeed $r_0$ invariant for
$\Lambda_{UV}^2 \stackrel{r_0}{\longrightarrow} - \Lambda_{UV}^2$.

The one-loop, renormalized, effective potential can be written (see, e.g.,~\cite{Pokorski:2000ruo}) as
\begin{widetext}
\begin{align}
	\!\!\!\!\!\!\!\!\!\!\!\!\! V_\text{eff}^\text{1-loop}(r^\mu) & =
	\delta m_0^2 r_0+(m_1^2+\delta m_1^2) r_2 + (m_{12}^2+\delta m_{12}^2) r_1 + (\lambda_1+\delta\lambda_1+\lambda_3+\delta\lambda_3)r_1^2 + 2(\lambda_1+\delta\lambda_1) r_2^2   \label{ren_pot1}  \\
	& + \frac{\Lambda_{\text{UV}}^2}{8\pi^2}(3 \lambda_1+\lambda_3) r_0  + \label{ren_pot2} \\
	& - \frac{\log\Lambda_{\text{UV}}^2}{32\pi^2}
	\left[ 
	4 m_1^2(3\lambda_1-\lambda_3)r_2 + 8 m_{12}^2 \lambda_3 r_1
	+ 4(9\lambda_1^2+6\lambda_1\lambda_3+5\lambda_3^2)r_1^2 + 8(9\lambda_1^2+\lambda_3^2)r_2^2 \right] + \label{ren_pot3} \\
	& + \text{finite~contributions,} \label{ren_pot4}
\end{align}
\end{widetext}
where $\delta \cdots$ denotes a corresponding counterterm.
Since during renormalization all dependence on the cutoff shall disappear, it is interesting to see how the $r_0$ symmetry
is realized in the renormalized potential.

Field-independent one-loop generated terms, even if divergent, have been dropped in (\ref{ren_pot2}) and (\ref{ren_pot3}).
Note that, since the cutoff $\Lambda_{\text{UV}}^2$ is odd under $r_0$, the counterterm $\propto \Lambda_{\text{UV}}^2 \cdot r_0$
is invariant, so that it can cancel the quadratic divergence in \eqref{ren_pot2} if
\beq
\delta m_0^2  = - \frac{(3 \lambda_1+\lambda_3)}{8\pi^2}\Lambda_{\text{UV}}^2\,.
\label{eq:delm0}
\eeq
The remaining counterterms in \eqref{ren_pot1} could be also adjusted to cancel divergent terms present in~\eqref{ren_pot3},
\begin{widetext}
\begin{align}
	\left.\frac{\dd V_\text{eff}^\text{1-loop}}{\dd r_2}\right|_{r^\mu=M}=m_1^2 & \hsp \Rightarrow \hsp \delta m_1^2  = +
	\frac{m_1^2(3\lambda_1-\lambda_3)}{8\pi^2}
	\log\left( \Lambda_{\text{UV}}^2 \right) \,, \\
	\left.\frac{\dd V_\text{eff}^\text{1-loop}}{\dd r_1}\right|_{r^\mu=M}=m_{12}^2 & \hsp \Rightarrow \hsp \delta m_{12}^2  = +
	\frac{m_{12}^2\lambda_3}{4\pi^2}
	\log\left( \Lambda_{\text{UV}}^2 \right)  \,,
\end{align}
for the quadratic terms, and for the quartic ones,
\begin{align}
	\left.\frac{\dd ^2V_\text{eff}^\text{1-loop}}{\dd r_2^2}\right|_{r^\mu=M}=4 \lambda_1 & \hsp \Rightarrow \hsp \delta \lambda_1 =
	+ \frac{i}{128 \pi} \frac{\sqrt{\Delta_0}}{M^2}(3\lambda_1+\lambda_3)
	+\frac{(9\lambda_1^2+\lambda_3^2)}{8\pi^2}
	\log\left( \Lambda_{\text{UV}}^2 \right)\,, \\ 
	\left.\frac{\dd^2V_\text{eff}^\text{1-loop}}{\dd r_1^2}\right|_{r^\mu=M}=2(\lambda_1+\lambda_3) & \hsp \Rightarrow \hsp \delta \lambda_3 =
	+ \frac{i}{128 \pi} \frac{\sqrt{\Delta_0}}{M^2}(3\lambda_1+\lambda_3)
	+\frac{\lambda_3 (3\lambda_1+2\lambda_3)}{4\pi^2}
	\log\left( \Lambda_{\text{UV}}^2 \right)\,, 
\end{align}
\end{widetext}
where $\Delta_0 \equiv \Delta|_{r^\mu=0}=(m_1^2)^2+(m_{12}^2)^2$. In order to regularize infrared (IR) singularities present at $r^\mu=0$,
the above renormalization conditions have been specified at $r^\mu=M$ with $M$ being a IR-regularization scale such that $r_1=r_2=M^2/\sqrt{2}$.
The counterterms have been chosen to contain only IR divergent ($M\to 0$) and/or the UV divergent ($\Lambda_{\text{UV}}\to \infty$) terms.
After dropping irrelevant field-independent terms (even if they are divergent), the renormalized one-loop effective potential reads as
\bea
V^{\text{1-loop}}_{\text{eff}}(r^\mu)&=&V^0(r_1,r_2)+
\frac{i}{64 \pi} \frac{\sqrt{\Delta_0}}{M^2}(3\lambda_1+\lambda_3)(r_1^2+r_2^2)\nonumber\\
&& + \sum_{i=1,2}\frac{M_i^4(r^\mu)}{64\pi}\left[\log M_i^2-\frac12 \right]+\cdots,
\label{eff_pot_cut_ren}
\eea
where the ellipsis denotes the terms hidden in \eqref{ren_pot4}. Since the curvature of the tree-level potential might be negative (e.g. in the vicinity of the origin), the one-loop effective potential $V^{\text{1-loop}}_{\text{eff}}$ might be a complex function~\cite{Weinberg:1987vp}.
Note also that in the scale-invariant limit $\Delta_0\to 0$ the imaginary part of the potential vanishes
in accordance with our intuition gained from a single massless real field model~\cite{Coleman:1973jx}. It is seen that the first two terms in \eqref{eff_pot_cut_ren} are trivially $r_0$ invariant, and since
$M_1^2  \stackrel{r_0}{\longleftrightarrow} - M_2^2$, the third term is invariant as well.




The ultimate verification of the quality of the toy model is to check if $\beta$ functions for squared mass terms which are forbidden by the $r_0$ symmetry
do vanish as they did in 2HDM.
Using generic results from \cite{Luo:2002ti} it is easy to show that for the potential without the $r_0$ symmetry~\eqref{r_0_def} imposed, the beta functions read as
\bea
	\beta_{m_1^2+m_2^2}&=&12(\lambda_1 m_1^2+\lambda_2 m_2^2)+2 \lambda_3 (m_1^2+m_2^2)\nonumber\\
	&&+6(\lambda_6+\lambda_7)m_{12}^2\\
	\beta_{m_{12}^2}&=&3(\lambda_6m_1^2+\lambda_7m_2^2)+4\lambda_3m_{12}^2,
\eea
where, for pedagogical reasons, here we have kept $m_2^2 \neq -m_1^2$ and $\lambda_{6,7}\neq 0$.
It is then clear that when the $r_0$ symmetry~\eqref{r_0_def} is imposed  $\beta_{m_1^2+m_2^2}$ vanishes.
This was also verified at two loops in~\cite{Haber:2025cbb}.
Therefore we conclude that the toy model constitutes the minimal model possessing the $r_0$ symmetry.
\section{Summary and conclusions}
\label{sec:conclusions}
In this paper we have discussed implications of the $r_0$ symmetry for the one-loop effective potential
within the 2HDM and the minimal (toy) model defined by~\eqref{pot2}. The novel aspect of the symmetry 
lies in the necessity to transform, via an imaginary scaling, coordinates together with
 fields. In this paper we showed that an imaginary scaling of scalar fields is accompanied by spacetime coordinates also being scaled by imaginary factors, so that invariance of the theory at tree level is achieved. We then showed that that procedure is consistent at one loop, and the effective potential
was shown to be invariant under the $r_0$ transformation. The one-loop effective potential has been calculated and renormalized.
For the invariance to hold the key property is that the imaginary scaling of spacetime coordinates implies an imaginary
scaling of 4-momenta, which in turn implies a nontrivial transformation of the cutoff, $\Lambda_{UV}$,  or renormalization, $\mu$, scales--namely
$\Lambda_{UV}^2 \stackrel{r_0}{\longrightarrow} - \Lambda_{UV}^2$ or $\mu^2 \stackrel{r_0}{\longrightarrow} - \mu^2$.

It is worth noticing that the invariance of beta functions, which contain coefficients of singular contributions to Green's functions, does not require imaginary scaling of coordinates. However, finite parts of loop corrections, like those that emerge in the effective potential, do need coordinate scaling. 

Relations between parameters of a theory invariant under the $r_0$ transformation 
are stable under RGE evolution to all orders. 
This is why we name the $r_0$ transformation a symmetry. However, this is not an ordinary symmetry
since the invariance requires complexification of real fields and coordinates. Nevertheless it   
is unquestionable that the parameter relations implied by the $r_0$ transformation are stable under RGE running.
This is the power and essence of the symmetry.

Symmetries in field theories imply the presence of relations between, {\it a~priori}, independent
parameters of a given model.
Sometimes they explain various fundamental facts, like, e.g., the masslessness of the photon, as a consequence of $U(1)$ gauge symmetry. 
Symmetries may also protect decays of dark matter candidates or, by forbidding certain mass terms, be relevant for the SM hierarchy problem. It is worth mentioning that field theories are applicable also in soft matter physics, statistical physics, and other fields; therefore the novel symmetries described in this work may prove to be useful in other contexts. Hence, proving the existence of hitherto unknown symmetries will broaden our understanding of bosonic field theories, and it might be relevant while looking for solutions of SM difficulties. It should be emphasized that within the 2HDM adopting the new transformations we obtain seven extra symmetries on top of the classical six ones.

\section*{Acknowledgments}
The authors thank Per Osland for his interest at the beginning of this project. B.G. thanks Apostolos Pilaftsis and Jochum van der Bij for inspiring remarks.
P.M.F. is supported
by \textit{Funda\c c\~ao para a Ci\^encia e a Tecnologia} (FCT)
through Contracts No.
UIDB/00618/2020, No. UIDP/00618/2020, No. CERN/FIS-PAR/0025/2021, and No. 2024.03328.CERN.
The work of B.G. is supported in part by the National Science Centre (Poland) as a research Project No. 2023/49/B/ST2/00856.

\section*{Data availability}
No data were created or analyzed in this study.

\bibliography{biblio}

\end{document}